\documentclass[12pt,preprint]{aastex}

\def\au{{\rm AU}} 

\def\masyr{{\rm mas}\,{\rm yr}^{-1}}

\def\max{{\rm max}}
\def\rel{{\rm rel}}
\def\e{{\rm E}}

\begin{document}
\title{Microlens Terrestrial Parallax Mass Measurements: A Rare Probe of Isolated Brown Dwarfs and Free-Floating Planets}

\author{Andrew Gould \& Jennifer C.\ Yee}
\affil{Department of Astronomy, Ohio State University,
140 W.\ 18th Ave., Columbus, OH 43210, USA; 
gould,jyee@astronomy.ohio-state.edu}

\begin{abstract}

Terrestrial microlens parallax is one of the very few methods that can
measure the mass and number density of isolated dark low-mass objects,
such as old free-floating planets and brown dwarfs.  Terrestrial microlens
parallax can be measured whenever a microlensing event differs substantially
as observed from two or more well-separated sites.  If the lens
also transits the source during the event, then its mass
can be measured.  We derive an analytic
expression for the expected rate of such events and then use this
to derive two important conclusions.  First the rate is directly
proportional to the number density of a given population, 
greatly favoring low-mass populations relative to their contribution to
the general microlensing
rate, which further scales as $M^{1/2}$ where $M$ is the lens mass.
Second, the rate rises sharply as one probes smaller source stars,
despite the fact that the probability of transit falls directly with
source size.  We propose  modifications
to current observing strategies that could yield a factor 100 increase in
sensitivity to these rare events.

\end{abstract}


\section{{Introduction}
\label{sec:intro}}

It is extremely difficult to detect and measure the mass of 
dark isolated objects and systems.  
The only known method is to detect the object in
a gravitational microlensing event and then to measure
two non-standard parameters:
the angular Einstein radius (on the plane of the sky) $\theta_\e$ 
and the projected Einstein radius (on the observer plane) $\tilde r_\e$.
The lens mass and distance are then given by \citep{gould92}
\begin{equation}
M = {\theta_\e\over\kappa \pi_\e},
\qquad
D_L = {{\rm AU}\over \pi_\rel + \pi_S}
\qquad
\kappa\equiv {4 G\over c^2\,{\rm AU}}\sim 8.1\,{{\rm mas}\over M_\odot}
\label{eqn:massidst}
\end{equation}
where $\pi_\rel\equiv (\au/D_L - \au/D_S)$ is the lens-source relative
parallax, $\pi_S$ is the source parallax, and
\begin{equation}
\pi_\rel = \theta_\e\pi_\e,
\qquad
\pi_\e \equiv {{\rm AU}\over \tilde r_\e}.
\label{eqn:kappadef}
\end{equation}
These require measurement of two independent 
higher-order effects (microlens-parallax effects and finite-source effects) 
each of which is only rarely measured individually.
Hence, mass measurements of dark lenses are quite rare.

An important exception to this rule is binary lensing, of which
planetary lensing may be considered a special case.  Binary lenses have
extended caustics, and when the microlensed source passes close to or
over one of these caustics (which it does in the majority of
recognized binary-lens events), then finite-source effects become important,
so that one can easily determine 
$\rho=\theta_*/\theta_\e$, the ratio of the angular size of the source to the
angular Einstein radius.  Using standard techniques \citep{yoo04},
one can then infer $\theta_*$ from the source position on the color-magnitude
diagram and so obtain $\theta_\e$.
It is still relatively rare that the second higher-order parameter $\pi_\e$
can be measured, but at least a concatenation of two rarities is not required.
Hence there are of order a dozen such measurements.

However, for isolated lenses, the caustic structure is simply a point
at the position of the lens.  Unless the limb of the source passes
directly over this point, $\rho$ cannot be measured 
photometrically.\footnote{It can in principle be measured astrometrically
\citep{hog95,miyamoto95,walker95}, but this requires a level of astrometric
precision that has not yet been achieved.}  The probability of such a chance 
alignment is simply $\rho$, and since 
$\rho=\theta_*/\theta_\e\sim \rm {\cal O}(\mu as)/{\cal O}(mas)\sim 10^{-3}$,
such events occur only a few times among the several thousand events
discovered each year.
Nevertheless, as \citet{gould97} pointed out, it is just these
extreme microlensing events (EMEs), with peak magnifications
$A_\max\sim \rho^{-1}\sim 10^3$ that could be susceptible to
a ``terrestrial parallax'' measurement of $\pi_\e$.

In principle, the ``microlens parallax'' $\pi_\e$ can be measured 
whenever observations are carried out from two or more locations 
within the Einstein ring projected onto the observer plane. This is because
the event appears different from the two locations, and the amount
of difference scales as $\tilde r_\e^{-1}$ (or linearly with $\pi_\e$).
In practice, however, events normally appear identical at different
locations on Earth because typically $\tilde r_\e\sim$ 1--10 AU, i.e.,
$\sim 10^{4.5}$ times larger than the distance between observatories.
For this reason, \citet{refsdal66} originally proposed that microlens
parallaxes be measured from a satellite in solar orbit.  \citet{gould92}
proposed an alternate method: making use of the Earth's orbital
motion to measure $\pi_\e$, but this requires events that remain
substantially magnified for a large fraction of a year, and these are
rare.  Moreover, very long events most often have large Einstein
radii, which reduces further the probability of the lens 
passing over the source limb.

\citet{hardy95} showed that given the steep magnification profiles
characteristic of binary-lens caustics, it would be possible to
distinguish the lightcurves even from two observatories on Earth.
Then \citet{holz96} pointed out that given enough photons, one
could in principle distinguish the lightcurves of the much smoother
point-lens events.  \citet{gould97} effectively combined these two ideas
by noting that during EMEs, the source passes very close to the
point-lens caustic (the very feature that permits a measurement of 
$\theta_\e$), thus also permitting a terrestrial parallax measurement of
$\pi_\e$, and so $M=\theta_\e/\kappa\pi_\e$.

Here we derive an analytic formula for the rate of terrestrial parallax
mass measurements and use this to draw several important conclusions.
We show that the actual number (2) of published
terrestrial parallax mass measurements is higher than predicted by
this formula and examine possible reasons for this.  Finally, we 
propose methods to greatly increase the rate of terrestrial parallax mass
measurements in the future.

\section{{Rate of Terrestrial Parallax Measurements}
\label{sec:rate}}

To measure the mass using terrestrial parallax,
four conditions are required.  First, the source size projected onto
the observer plane must be $\rho\tilde r_\e \la 50\,R_\oplus$.  Otherwise
the difference in magnifications ${\cal O}((\rho\tilde r_\e)^{-1})$
will be less than a few percent, making robust measurement difficult. 
Since $\rho=\theta_*/\theta_\e$ and $\pi_\rel = \theta_\e\pi_\e$, this implies 
\begin{equation}
\pi_\rel= \theta_\e\pi_\e =
{\au\over\rho\tilde r_\e}\theta_* \ga 0.28\,{\rm mas}
{\theta_*\over 0.6\,\mu\rm as}
\label{eqn:rho}
\end{equation}
Hence, for typical microlensed sources, $\theta_*\ga 0.6\,\mu$as, the
lens should be closer than $D_\max = 2.5\,$kpc.

Second, the mass measurement requires that the lens transit the source.
The rate of such events per star that also satisfy the first condition is
\begin{equation}
\Gamma = 2\langle\mu\rangle\theta_*\int_0^{D_\max} d D_L D_L^2 n(D_L)
= 1.6\,{\rm Gyr}^{-1}
\biggl(
{\langle\mu\rangle\over 10\,{\rm mas/yr}}
\biggr)
\biggl(
{\theta_*\over 0.6\,\mu{\rm as}}
\biggr)
\biggl({D_\max\over 2.5\,\rm kpc}\biggr)^3
\biggl(
{\langle n \rangle\over 1\,\rm pc^{-3}}
\biggr)
\label{eqn:rate}
\end{equation}
where $n(D_L)$ is the local number density of lenses, $\langle n\rangle$
is its mean over the volume, and $\langle\mu\rangle$ is the mean
lens-source relative proper motion.  Note in particular that the rate
depends only on the number density of lenses, not on their mass.  
This favors brown dwarfs and free-floating planets over stars
because they are more common \citep{sumi11}.

Third, the peak of the event must be simultaneously
observable from two sites that are
separated by a substantial fraction of $R_\oplus$.  This
imposes three constraints.  First, if the observatories are too close,
then they will lack sufficient baseline for a measurement.  Second,
if they are too far apart (such as Chile and New Zealand) then their
observing windows will rarely overlap.  Third, the event must occur
within the 3-4 months of the peak of the observing season or simultaneous
observation from well-separated observatories is not possible.
An exception would be pairing northern observatories (each of which
has an extremely short observing window) with southern observatories.
It should be noted both published terrestrial-parallax mass
measurements (described below)
have in fact combined northern and southern observations.

Finally, obviously, the event must actually be observed.  Aggressive
observation of high-magnification events has been ongoing since 2004,
with roughly half of all cataloged high-mag $A_{\rm max}>200$ events effectively
covered \citep{gould10}.  If we assume that $n\sim 1\,\rm pc^{-1}$ stars,
brown dwarfs, and free-floating planets locally \citep{sumi11},
that the target season is 1/4 of the year, that about 1/10 of
events peak when they can be simultaneous observed from widely separated sites,
and that the
surveys effectively monitor $N\sim 5\times 10^8$ sources (including those
blended with other stars), we would have
expected $(1/4)(1/10)(1/2)\Gamma N T= 0.1$ terrestrial parallax mass 
measurements, where $T=10\,$yr is the duration of the search to date.
Hence, the probability of having two such
measurements is about $5\times 10^{-3}$, which is small enough that
a closer examination of the detections is warranted.

\section{{Comparison to Observations}
\label{sec:comp}}

To date, there have been two published terrestrial parallax mass measurements,
OGLE-2007-BLG-224 \citep{ob07224} and OGLE-2008-BLG-279 \citep{ob08279}.
The key features of these events are compared in Table 1.


\begin{table}
\caption{\label{tab:parameters} \sc Events with Terrestrial Parallax}
\vskip 1em
\begin{tabular}{@{\extracolsep{0pt}}llrrrrrr}
\hline
\hline
Name & {$M$} & {$D_L$} & {$\mu$} & {$\theta_*$} & {$\rho\tilde r_\e$}& 
{$t_\e$}& {$A_{\rm max}$} \\ \hline
 & {($M_\odot$)} & (kpc) & (mas/yr) & ($\mu$as) & ($R_\oplus$) &(day)& \\ \hline
\hline
OGLE-2007-BLG-224 & 0.056 & 0.5 & 48.0  & 0.77 & 10 & 106& 2400\\ \hline
OGLE-2008-BLG-279 & 0.64  & 4.0 & 2.7 & 0.54 & 100 &    7& 1600\\ \hline
\end{tabular}
\end{table}

The only thing that is ``typical'' about these events relative to the
fiducial numbers in Equation~(\ref{eqn:rate}) is that the source sizes
are typical for observed microlensing events, $\theta_*\sim 0.6\,\mu$as.
Regarding peculiar features, let us first examine OGLE-2008-BLG-279.
Terrestrial parallax was measured despite the fact that 
$\rho\tilde r_\e=100\,R_\oplus$, twice the value suggested above.
As a direct consequence of this fact and Equation~(\ref{eqn:rho}),
terrestrial parallax was detected 
at $\pi_\rel\sim 0.13\,\masyr$, i.e., at a distance
1.6 times larger than estimated in Equation~(\ref{eqn:rate}), which
encloses a 4-times larger volume.  Although not immediately obvious,
this detection was made possible by the relatively slow proper motion
and relatively high lens mass.  Together, these resulted in an exceptionally
long Einstein timescale $t_\e=106\,$days.  Hence, despite the extremely
high magnification, the effective timescale $t_{\rm eff}=1.6\,$hr was long
enough to enable very dense observations over the peak from multiple
observatories, which in turn permitted more precise measurement of subtle
effects.  See Figure~1 of \citet{ob08279}.

OGLE-2007-BLG-224 by contrast was detected at only $D_L=0.5\,$kpc, i.e.,
within a volume that is 125 times smaller than envisaged by 
Equation~(\ref{eqn:rate}).  This proximity is partially responsible
for the high proper motion, which somewhat compensates
for the reduced volume.  However, the primary reason for the high proper
motion is that the lens is in the thick disk, whose number density is
much lower than the thin-disk normalization of Equation~(\ref{eqn:rate}).
Due to the low mass and high proper motion, OGLE-2007-BLG-224 had a very
short $t_\e=7\,$days, 
so that given the high magnification, the effective timescale
was only $t_{\rm eff}\sim 4\,$min, by far the shortest ever recorded.
This made it extremely difficult to organize and take observations over
the peak, but the exceptionally small $\rho\tilde r_\e=10\,R_\oplus$
meant that a robust terrestrial parallax measurement was still very
feasible.

In brief, neither of these two events ``fits the mold'' sketched
by Equation~(\ref{eqn:rate}).  Whether these discrepant features
are connected with the higher-than-expected event rate cannot be
assessed without more events of this type.

\section{{Increasing the Rate of Terrestrial Parallax Mass Measurements}
\label{sec:increase}}

Because terrestrial-parallax mass measurements are a unique probe
of low-mass isolated objects, it is worth some thought as to how
to increase their rate.  We present three ideas that, together, could
improve the effective sensitivity by two orders of magnitude.
First, by aggressive monitoring of ongoing microlensing surveys,
it should be possible to recognize many more faint-source high-magnification
events in real time.  At present, only the MOA collaboration even
attempts to recognize ``new events'' arising from uncatalogued sources
in real time, and MOA observing conditions are far less ideal
than those in Chile or Africa where other existing and planned surveys
are located.  Such faint-star sources are extremely important because
the total rate per star scales as
\begin{equation}
\Gamma \propto \theta_* D_\max^3\sim \theta_*^{-2}
\biggl[1+ {0.4\over \theta_*/0.6\,\mu\rm as}\biggr]^{-3},
\label{eqn:rate_full}
\end{equation}
where we have used $D_L = \au[\theta_*(\au/\rho\tilde r_\e) +\pi_S]^{-1}$
and assumed $\rho\tilde r_\e < 50\,R_\oplus$ and $\pi_S=125\,\mu$as.
Hence, physically smaller sources each have a much higher rate, and
there are more small stars than big stars.  Because terrestrial-parallax
mass measurements already {\it require} high-magnification, these
intrinsically faint sources will still yield high signal-to-noise ratio
measurements.  Now, in principle, even if these are not announced
in real time, they may still be simultaneously monitored from two
continents by routine survey observations.  
However, as discussed in Section~\ref{sec:rate}, this is
relatively unlikely.

A second suggestion, then, is the addition of many northern
telescopes to the network of followup observatories.  As mentioned
above, each such observatory would have a very narrow window and so
a very limited number of high-magnification events that it could
monitor.  This would be an advantage in the sense that it would
require a limited commitment.  By the same token, a large number
of such observatories would be needed to effectively cover the
24-hour day.  But with such coverage, the number of monitored events
could be increased by a factor 10 by creating a $\sim 1\,R_\oplus$
north-south baseline for almost all events in place of the
current $\sim 1\,R_\oplus$ east-west baseline that exists for a small fraction
of events.

Finally, at present there are a very large number of microlensing
events discovered by survey teams in low-cadence fields whose
nature as high-magnification cannot be effectively predicted based on these
sparse data.  These events could be monitored by a network of
``robotically intelligent''
narrow angle telescopes, with feedback loops aimed at acquiring
enough data to adequately predict high magnification.  The same
feedback loops could then enable these telescopes 
to undertake the several-site intensive monitoring
needed for terrestrial parallax measurements.  
RoboNet \citep{tsapras09} is an example of such a network,
which is presently under construction and is in partial operation.
It uses web-PLOP \citep{snodgrass08} to compile an optimal list
of targets and SIGNALMEN \citep{dominik07} and ARTEMiS \citep{dominik08} 
to evaluate possible lightcurve anomalies and redirect observations.  
While prediction and multi-site observations of terrestrial parallax
events is not presently a goal of this network, it could be adapted
to this purpose without major modifications.

\section{{Conclusions}
\label{sec:conclude}}

Terrestrial parallax is one of the very few methods of measuring the
mass and distance of isolated, dark, low-mass objects.  We have
shown that the rate of such events is directly proportional the
{\it number} of target objects, which greatly favors brown dwarfs and
planets since these probably account for a majority of the number of
all lenses, but a small fraction of the event rate.

To date, only two terrestrial parallax mass measurements have been made,
but this already greatly exceeds the the number expected based on the
estimate given by Equation~(\ref{eqn:rate}).  The reason for the discrepancy
is not understood.  It may be connected with the ``unusual'' character
of these two events, which we detailed in Section~\ref{sec:comp}, or it
may be a $\sim 5\times 10^{-3}$ statistical fluctuation.

The rate of terrestrial parallax mass measurements could be increased
by a factor 10 simply by monitoring the relatively rare candidate
events from many more sites, particularly in the northern hemisphere.
A further increase of several-to-ten could be achieved by aggressively
identifying intrinsically faint sources for possible high magnification,
since smaller sources are much more likely to yield terrestrial parallax
mass measurements according to Equation~(\ref{eqn:rate_full}).  Finally,
a large fraction of high-magnification events are currently going
unharvested in low-cadence survey fields, which could be rectified
by a network of robotic telescopes such as RoboNet.

\acknowledgments

This work was supported by NSF grant AST 1103471.
J.C. Yee is supported by a Distinguished University Fellowship
from The Ohio State University.


\end{document}